\def\N{\mathbb N}
\def\Z{\mathbb Z}
\def\R{\mathbb R}
\def\d{\mathrm d}
\def\C{\mathbb C}
\def\Q{\mathbb Q}
\def\A{\mathcal A}

\def\pfz{\begin{proof}}
\def\pfk{\end{proof}}

\documentclass[preprint,12pt]{elsarticle}
\usepackage{amssymb,amsmath,amsthm}
\usepackage{rotating}
\usepackage{algorithmic,algorithm}
\usepackage{lscape}
\newtheorem{lem}{Lemma}
\newtheorem{thm}[lem]{Theorem}
\newtheorem{prop}[lem]{Proposition}
\newtheorem{coro}[lem]{Corollary}
\newtheorem{de}[lem]{Definition}
\newtheorem{ex}[lem]{Example}
\newtheorem{pozn}[lem]{Remark}
\newtheorem{claim}[lem]{Claim}

\journal{Theoretical Computer Science}

\begin{document}
\begin{frontmatter}
%%%%%%%%%%%%%%%%%%%%%%%%%%%%%%%%%%%%%%%%%%%%%%%%%%%%%%%%%%%%%%%%%%%%%%%%%%
\title{Number representation using generalized $(-\beta)$-transformation}
\author{D. Dombek, Z. Mas\'akov\'a\corref{e-mail: zuzana.masakova@fjfi.cvut.cz}, E. Pelantov\'a}%\\[1mm]
%{\normalsize Department of Mathematics FNSPE, Czech Technical University in Prague}\\
%{\normalsize Trojanova 13, 120 00 Praha 2, Czech Republic}}

\address{Department of Mathematics FNSPE, Czech Technical University in Prague\\
Trojanova 13, 120 00 Praha 2, Czech Republic}

\begin{abstract}
We study non-standard number systems with negative base $-\beta$. Instead of the Ito-Sadahiro definition, based
on the transformation $T_{-\beta}$ of the interval $\big[-\frac{\beta}{\beta+1},\frac{1}{\beta+1}\big)$ into itself, we suggest a
generalization using an interval $[l,l+1)$ with $l\in(-1,0]$. Such generalization may eliminate certain disadvantages of the Ito-Sadahiro system. We focus on the description of admissible digit strings and their periodicity.
\end{abstract}


\end{frontmatter}
%%%%%%%%%%%%%%%%%%%%%%%%%%%%%%%%%%%%%%%%%%%%%%%%%%%%%%%%%%%%%%%%%%%%%%%%%%
\section{Introduction}

In 2008, Ito and Sadahiro~\cite{ItoSadahiro}, introduced a numeration system with negative base $-\beta$, where $\beta>1$. It has been shown that every number
$x\in I=\Big[-\frac{\beta}{\beta+1},\frac1{\beta+1}\Big)$ can be written in the form
$$
x=\sum_{i=1}^\infty \frac{x_i}{(-\beta)^i}\,,\qquad x_i\in\{0,1,\dots,\lfloor\beta\rfloor\}\,.
$$
The string of digits $\d(x)=x_1x_2x_3\cdots$, which we call here Ito-Sadahiro expansion of $x$, can be obtained by the transformation $T:I\mapsto I$ defined by
$$
T(x) = -\beta x-\Big\lfloor -\beta x + \frac{\beta}{\beta+1}\Big\rfloor\,.
$$
The digits of the string $\d(x)$ are then given by
$$
x_i=\Big\lfloor-\beta T^{i-1}(x) + \frac{\beta}{\beta+1}\Big\rfloor\,,
$$
i.e. $x_i$ lie in the alphabet $\A=\{0,1,\dots,\lfloor\beta\rfloor\}$

The natural ordering of reals in the interval $I$ is preserved by Ito-Sadahiro expansion, when considering the alternate order $\preceq_{\text{alt}}$ on the set of digit strings over an ordered alphabet $\A$. We say that $x_1x_2x_3\cdots \prec_{\text{alt}}
y_1y_2y_3\cdots$ if the first non-zero element of the sequence $(-1)^i(y_i-x_i)$, $i\geq 1$, is positive.

The Ito-Sadahiro numeration system shares many properties of numeration with positive (in general non-integer) base considered by R\'enyi~\cite{Renyi} and then by many others from different points of view. %For an overview see ... XXXXXXXXXX~\cite{Parry,schmidt,FruSo}.
The main advantage of considering negative base is the possibility of representing every real number with the same set of digits without the use of sign. On the other hand, many aspects of number system with negative base are different and essentially more complicated, as an example, let us mention
the structure of $(-\beta)$-integers which is studied in~\cite{ADMP} and~\cite{Steiner}.

Similarly as in case of number representation using a positive base,
not every string of digits from the alphabet $\A=\{0,1,\dots,\lfloor\beta\rfloor\}$ appears as the expansion $\d(x)$ of some $x\in I$.
Ito and Sadahiro~\cite{ItoSadahiro} have shown that a necessary and sufficient condition so that $x_1x_2x_3\cdots$ be equal to $\d(x)$ for some $x$
is that
\begin{equation}\label{eq:*}
\d\Big(\frac{-\beta}{\beta+1}\Big) \preceq_{\text{alt}} x_ix_{i+1}x_{i+2}\cdots \prec_{\text{alt}} \d^*\Big(\frac{1}{\beta+1}\Big) \qquad\text{for all $i=1,2,3,\dots$}\,,
\end{equation}
where
$$
\d^*\Big(\frac{1}{\beta+1}\Big)=\lim_{\varepsilon\to 0+} \d\Big(\frac{1}{\beta+1}-\varepsilon\Big)\,.
$$
The upper and lower bounds deciding about admissibility of a digit string $x_1x_2x_3\cdots$ as the Ito-Sadahiro expansion are closely related.
If the lower bound is a purely periodic digit string with odd period-length, i.e. $\d\big(-\frac{\beta}{\beta+1}\big) = (d_1d_2\cdots d_{2q+1})^\omega$, where $w^\omega$ stands for infinite repetition of the string $w$,
then $\d^*\big(\frac{1}{\beta+1}\big) = \big(0d_1d_2\cdots d_{2q}(d_{2q+1}-1)\big)^\omega$. Otherwise, $\d^*\big(\frac{1}{\beta+1}\big) = 0\d\big(-\frac{\beta}{\beta+1}\big)$.
Such a close relation of $\d\big(-\frac{\beta}{\beta+1}\big)$ and $\d^*\big(\frac{1}{\beta+1}\big) $ allowed to show that the $(-\beta)$-shift is sofic if and only if the lower bound is eventually periodic~\cite{ItoSadahiro}. Frougny and Lai~\cite{ChiaraFrougny} have shown that for $\beta$ Pisot number ever element of $\Q(\beta)\cap I$ has eventually periodic Ito-Sadahiro expansion, and so for such $\beta$, the $(-\beta)$-shift is sofic.

Using base $-\beta$, Ito and Sadahiro have represented only numbers within the interval $I$. Since for every $x\in\R$ there exists $k\in\Z$ such that $\frac{x}{(-\beta)^k}\in I$, every $x\in\R $ can be written in the form
$$
x=x_k(-\beta)^k + x_{k-1}(-\beta)^{k-1}+ x_{k-2}(-\beta)^{k-2}+\cdots\,,
$$
where any suffix of the digit string $x_kx_{k-1}x_{k-2}\cdots$ verifies the condition~\eqref{eq:*}. The use of negative base thus allows one to represent every real number  using the same set of digits $\{0,1,\dots,\lfloor\beta\rfloor\}$. A disadvantage is that the representation of $x$ has no obvious relation to the representation of $-x$. More serious from the practical point of view is, however, the fact that the digit strings for representation of $x$ and $\frac{x}{-\beta}$ can substantially differ (see discussion in Example~\ref{ex:itosadahiroproblem}). Recall that in a number system with positive base $\alpha$, the representation of $\frac{x}{\alpha}$ is always obtained from the representation of $x$ by shifting the fractional point. Among the disadvantages from the arithmetical point of view, we can mention that for some bases $-\beta$, the number zero is the only element of $I$ for which the Ito-Sadahiro expansion has only finitely many non-zero digits. In the positive base number systems such phenomenon does not occur.

In their paper, Ito and Sadahiro do not explain the reasons for choosing for representation the transformation $T$ with domain $I=\Big[-\frac{\beta}{\beta+1},\frac1{\beta+1}\Big)$. The aim of this paper is to discuss the influence of the choice of the domain on the properties of the resulting number system. Similar study for the positive base with focus on the dynamical aspects and tilings has been performed in~\cite{KalleSteiner}.

We consider the following generalization of the Ito-Sadahiro expansion.

\begin{de}
For a given real number $\beta>1$ and $l\in(-1,0]$ we define the mapping $T:I\mapsto I$ with domain $I:=[l,l+1)$, by the prescription
\begin{equation}\label{eq:cislo1}
T(x):= -\beta x - \lfloor -\beta x-l\rfloor\,,\qquad\text{for $x\in I$.}
\end{equation}
The mapping $T$ will be called the $(-\beta,l)$-transformation. To every $x\in I$ we associate an infinite  string of integer digits
$\d(x)=x_1x_2x_3\cdots$ by
\begin{equation}\label{eq:cislo2}
x_i:= \lfloor -\beta T^{i-1}(x)-l\rfloor\,,\qquad\text{for any $i=1,2,3,\dots$.}
\end{equation}
The mapping $\d:I\mapsto \Z^\N$ will be called the $(-\beta,l)$-expansion. The string $\d(x)$ is the $(-\beta,l)$-expansion of $x\in I$.
\end{de}

The paper is organized in the following way. In Section~\ref{sec:kladna} we recall the properties of number systems with positive real base.
In Section~\ref{sec:repre} we study number representation in a system with negative base from a general point of view, based on theorem of Thurston~\cite{thurston}, see Theorem~\ref{thm:thurston}. We explain in what sense our choice of transformation $T$ is general.
In section~\ref{sec:transformation} we focus on the properties of the $(-\beta,l)$-transformation and study how the choice of its domain influences
certain aspects of the corresponding number system. In Section~\ref{sec:admis} we characterize for fixed $\beta$ and $l$ the family of strings in $\Z^\N$ which are $(-\beta,l)$-expansions of some $x$. Analogously to the case of Ito-Sadahiro expansions, we use the alternate order on $\Z^\N$ and two reference digit strings denoted by $\d(l)$ and $\d^*(r)$. In Section~\ref{sec:digitstrings} we show  which property of the transformation $T$ connects these two strings. We also describe some of their properties and discuss the question which digit strings may play the role of the reference strings $\d(l)$ and $\d^*(r)$ for some $\beta$ and $l$. Finally, in Section~\ref{sec:periodic} we prove that periodicity of $(-\beta,l)$-expansions relates to the notion of Pisot and Salem numbers, as it is the case of positive base number systems.

%%%%%%%%%%%%%%%%%%%%%%%%%%%%%%%%%%%%%%%%%%%%%%%%%%%%%%%%%%%%%%%%%%%%%%%%%%%%
\section{Representing reals using a positive base}\label{sec:kladna}

Let us briefly recall several facts about numeration in positive base number systems for which we try to find analogues in the case of systems with negative base associated with $(-\beta,l)$-transformation. For a real number $\beta>1$, R\'enyi in~\cite{Renyi} has considered the transformation $T_\beta(x):=\beta x - \lfloor\beta x \rfloor$ of the interval $[0,1)$. One can represent every number $x\in[0,1)$ in the form
$$
x=\sum_{i=1}^\infty \frac{x_i}{\beta^i}\,,\qquad x_i\in\big\{0,1,\dots,\lceil\beta\rceil-1\big\}\,.
$$
where the digits $x_i$, $i\geq 1$ are obtained by
$$
x_i=\Big\lfloor\beta T_\beta^{i-1}(x) \Big\rfloor\,.
$$
The string of these digits is denoted $\d_\beta(x)=x_1x_2x_3\cdots$ and called the $\beta$-expansion of $x$. One can define the $\beta$-expansion
of every positive real number $x$ by dividing $x$ by a suitable power of $\beta$ so that $\frac{x}{\beta^{k}}\in[0,1)$, finding the $\beta$-expansion $\d_\beta\big(\frac{x}{\beta^{k}}\big)$ and shifting the fractional point.

Not every infinite string with digits in $\big\{0,1,\dots,\lfloor\beta\rfloor\big\}$ is admissible as a $\beta$-expansion of some $x\in[0,1)$. The characterization of the admissible digit strings
uses the lexicographic order $\preceq_{\hbox{\tiny lex}}$ and the infinite R\'enyi $\beta$-expansion of 1, namely the string
$$
\d^*_\beta(1) = \lim_{\varepsilon\to0+} \d_\beta(1-\varepsilon)\,.
$$
Here we make use of the fact that the space $\A^\N$ of infinite words over any finite alphabet $\A$ is compact with respect to the product topology.
The limit of a sequence of digit strings which have longer and longer common prefixes thus can be defined.
The necessary and sufficient condition for admissibility of digit strings was formulated by Parry~\cite{Parry}.

\begin{thm}[Parry]\label{t:parry}
Let $\beta>1$. The string $x_1x_2x_3\cdots$ of integer digits is a $\beta$-expansion of some $x\in[0,1)$
if and only if every suffix $x_ix_{i+1}x_{i+2}\cdots$ of $x_1x_2x_3\cdots$ satisfies
\begin{equation}\label{eq:admisplus}
0^\omega\preceq_{\hbox{\tiny lex}} x_ix_{i+1}x_{i+2}\cdots \prec_{\hbox{\tiny lex}} d_\beta^*(1)\,.
\end{equation}
\end{thm}

As a consequence of the above theorem, Parry has also given a criterion for sequences of digit strings which can play role of the R\'enyi $\beta$-expansion of 1 for some $\beta$. In fact, a sequence of integers $x_1x_2x_3\cdots$ is equal to $\d^*_\beta(1)$ for some $\beta>1$ if and only if every suffix $x_ix_{i+1}x_{i+2}\cdots$ satisfies
$$
0^\omega \preceq_{\hbox{\tiny lex}} x_ix_{i+1}x_{i+2}\cdots \preceq_{\hbox{\tiny lex}} x_1x_2x_3\cdots\,.
$$

For a fixed $\beta>1$, the $\beta$-shift is the closure of the set of infinite sequences appearing as $\beta$-expansions of numbers in $[0,1)$, which is a shift-invariant subspace of the space of all infinite sequences over the same alphabet. Such a dynamical system is sofic, if the set of its finite factors is recognized by a finite automaton. This is equivalent to saying that the reference string $\d_\beta^*(1)$, which is plays a crucial role in Theorem~\ref{t:parry}, is eventually periodic. The base $\beta$ leading to a sofic $\beta$-shift is called a Parry number.
Here, the number $1$ is expressed as a power series $1=\sum_{i=1}^\infty \frac{x_i}{\beta^i}$, where the integer coefficients $x_i$ form an eventually periodic sequence, then $\beta$ is a root of a monic polynomial in $\Z[X]$, and so every Parry number is an algebraic integer.

When studying the positive base number systems from the arithmetical point of view, one is mainly interested in the set ${\rm Fin(\beta)}$ of numbers with finite $\beta$-expansions, in particular, in the set of numbers $x\in[0,1)$ such that $\d_\beta(x)$ ends in $0^\omega$. This set is always infinite. One can also study which numbers have eventually periodic expansion $\d_\beta(x)$. Two classes of algebraic numbers prove to be of particular interest, namely Pisot and Salem numbers. Pisot numbers are algebraic integers $>1$ whose all conjugates are in modulus $<1$, Salem numbers are algebraic integers whose all conjugates are in modulus $\leq 1$, with at least one unimodular conjugate.
Being a Pisot number turns out to be a necessary condition for the base so that the set ${\rm Fin}(\beta)$ has a ring structure~\cite{FruSo}, it allows to construct fractal tilings~\cite{arnouxito}, etc. It is also known~\cite{Parry} that all Pisot numbers are Parry numbers. The question whether Salem
numbers are Parry numbers has been only solved for Salem numbers of small degree.
Schmidt~\cite{schmidt} shows that Pisot numbers share in some sense the properties of natural numbers, since for Pisot bases $\beta$ the set of numbers in $[0,1)$ with eventually periodic $\beta$-expansions is equal to $\Q(\beta)\cap[0,1)$. On the other hand, if every $x\in\Q(\beta)\cap[0,1)$ has eventually periodic $\d_\beta(x)$ then $\beta$ is either Pisot or Salem.

%%%%%%%%%%%%%%%%%%%%%%%%%%%%%%%%%%%%%%%%%%%%%%%%%%%%%%%%%%%%%%%%%%%%%%%%%%%%
\section{Representing reals using a negative base}\label{sec:repre}

Let us approach the question of representing numbers using powers of a basis from a more general point of view.
Thurston~\cite{thurston} stated the following simple theorem.

\begin{thm}[Thurston]\label{thm:thurston}
Given $\alpha\in\C$, $|\alpha|>1$, a finite alphabet $\A\subset\C$ and a bounded set $V\subset\C$ such that
\begin{equation}\label{eq:cislo3}
\alpha V\subset \bigcup_{a\in\A}(V+a)\,.
\end{equation}
Then for every $z\in V$ there exists a sequence $a_1a_2a_3\cdots \in \A^\N$ such that
\begin{equation}\label{eq:cislo4}
z= \sum_{i=1}^\infty\frac{a_i}{\alpha^i}\,.
\end{equation}
If, moreover, the point 0 lies in the interior of $V$, then every $z\in\C$ can be written in the form
$$
z= b_k\alpha^k+b_{k-1}\alpha^{k-1}+b_{k-2}\alpha^{k-2}+\cdots \qquad\text{for some $k\in\Z$ and $b_i\in\A$.}
$$
\end{thm}

The proof of this theorem is simple and constructive, we give it here for the purpose of discussion of uniqueness of the
sequence $a_1a_2a_3\cdots$ dependently on the choice of the set $V$ and the alphabet $\A$.

\pfz
Consider $z\in V$. According to~\eqref{eq:cislo3}, there exist $z_1\in V$ and $a_1\in\A$ such that $\alpha z = z_1+a_1$, i.e.
$z=\frac{a_1}{\alpha}+\frac{z_1}{\alpha}$. Again, by~\eqref{eq:cislo3}, there exist $z_2\in V$ and $a_2\in\A$ such that $\alpha z_1 = z_2+a_2$, i.e.
$z=\frac{a_1}{\alpha}+\frac{a_2}{\alpha^2}+\frac{z_2}{\alpha^2}$. This procedure can be repeated, so that after the $n$-th step we obtain
$$
z=\frac{a_1}{\alpha}+\frac{a_2}{\alpha^2}+\cdots + \frac{a_n}{\alpha^n}+\frac{z_n}{\alpha^n}\,,\qquad\text{where $z_n\in V$ and $a_1,\dots,a_n\in\A$}.
$$
As $V$ is bounded and $|\alpha|>1$, we have easily $\lim\limits_{n\to\infty}\frac{z_n}{\alpha^n}=0$, which proves the first statement
of the theorem. The second statement follows easily from the following fact. Since $0$ belongs to the interior of $V$, one can find for every $z\in\C$
an exponent $k\in\Z$ such that $\frac{z}{\alpha^k}\in V$. The first statement ensures existence of a representation $\frac{z}{\alpha^k}=\sum_{i=1}^\infty\frac{a_i}{\alpha^i}$.
\pfk

We shall use the above method for representing real numbers $x$ by powers of a negative base $\alpha=-\beta$, $\beta>1$. We will, moreover,
require the following:
\begin{enumerate}
 \item The region $V$ is a bounded interval $I\subset\R$;
 \item the expression $x=\sum_{i=1}^\infty\frac{a_i}{(-\beta)^i}$ from the
 theorem is unique for every $x\in V$;
 \item the alphabet $\A\subset\R$ is minimal in the sense that for every $a\in\A$
 there is an $x\in\R$ in whose representation $a_1a_2a_3\cdots$ the
 letter $a$ appears at least once.
\end{enumerate}

Requirement 2.\ implies that $-\beta I\subset \bigcup_{a\in\A}(I+a)$, where the union is disjoint. By requirement 1.\
the set $-\beta I$ is an interval, which together with the disjointness means that $I$ is a semi-closed interval and, moreover, the intervals
$I+a$, $a\in\A$, concur without gaps and overlaps. Denoting $m=\min\A$, we obtain using requirement 3.\ that $\A=\{m,m+|I|,m+2|I|,\dots,m+(\#\A-1)|I|\}$, where $|I|$ denotes the length of the interval $I$. Since~\eqref{eq:cislo3}
remains valid when scaling both $V$ and $\A$ by the same fixed factor, we can, without loss of generality, set $|I|=1$ and
$I=[l,r)$ for some $l\in\R$, $r=l+1$. In such a case, the alphabet is of the form $\A=\{m+k\mid k=0,1,\dots,\#\A-1\}$.
Imposing another requirement,

\begin{enumerate}
\setcounter{enumi}{3}
\item $\A\subset\Z$,
\end{enumerate}

\noindent
the alphabet $\A$ becomes a finite set of consecutive integers.

Now one can provide a simple prescription which to a given $x$ assigns the first digit $a_1$ and the remainder $z_1$,
$$
-\beta x\in[l,r)+a_1 \quad\Leftrightarrow\quad l+a_1 \leq -\beta x < r+a_1 \quad\Leftrightarrow\quad
-\beta x -r < a_1 \leq -\beta x -l\,.
$$
Since $a_1\in\Z$, we obtain $a_1=\lfloor -\beta x - l\rfloor$ and $z_1 = -\beta x -a_1 = -\beta x - \lfloor -\beta x - l\rfloor$.
Note that $z_1=T(x)$, where $T$ is the transformation defined in~\eqref{eq:cislo1} and $a_1=x_1$ from~\eqref{eq:cislo2}.
The digits take values in the alphabet
\begin{equation}\label{eq:abeceda}
{\mathcal A}_{-\beta,l}:=\Big\{ \big\lfloor -l(\beta+1)-\beta \big\rfloor, \dots, \big\lfloor -l(\beta+1) \big\rfloor \Big\}\,.
\end{equation}
which depends on $l$ and $\beta$. Thus for any $\beta$ and $l$, the digits in the numeration system form a bounded set of
$P$ consecutive integers where $P=\lfloor\beta\rfloor+1$ or $P=\lfloor\beta\rfloor+2$.

One may impose some further natural requirements on the number system:
\begin{itemize}
\item That the digits can take the value 0. For that, we need that $-l(\beta+1)-\beta< 1$ and $-l(\beta+1)\geq 0$,
which is equivalent to $-1< l \leq 0$, i.e.\ $0\in[l,r)$.

\item That the expansion of 0 is equal to $\d(0)=000\cdots = 0^\omega$. This is satisfied if $0$ is a fixed point of the
transformation $T$, for which we need $\lfloor -l\rfloor = 0=T(0)$, which again gives the condition $0\in[l,r)$.

\item That the numeration system can be extended to represent not only numbers from $[l,r)$ but all reals. Here we need that
every real number can be multiplied by an integer power of $-\beta$ so that it falls within the interval $[l,r)$. Again, one gets the condition
$0\in[l,r)$.
\end{itemize}

The above items justify our choice of considering $-1<l\leq 0$, which we impose throughout the paper. Note that
one can derive by simple algebraic manipulation that
the digits in the alphabet~\eqref{eq:abeceda} are then bounded by $\lceil\beta\rceil$ in modulus.

%%%%%%%%%%%%%%%%%%%%%%%%%%%%%%%%%%%%%%%%%%%%%%%%%%%%%%%%%%%%%%%%%%%%%%%%%%%%%%%%%%%%%%%%%%%%%%%%%%%%%%%%%
\section{The $\boldsymbol{(-\beta,l)}$-transformation $\boldsymbol{T}$}\label{sec:transformation}

Let us now see how the choice of the domain $[l,l+1)=[l,r)$ of the transformation $T$ influences
the properties of the resulting numeration system, or, on the other hand,
how further requirements on the numeration system limit the choice of $l$.

One may prefer that the digits are all non-negative, as it is the case of R\'enyi expansions with positive base. Such a condition is equivalent to
the requirement $-l(\beta+1)-\beta\geq 0$, which implies $l\geq -\frac{\beta}{\beta+1}$.
When, moreover, one requires the digits to be smaller or equal to $\beta$, we must ask for $-l(\beta+1)<\lfloor\beta\rfloor+1$, i.e.\
$l<-\frac{\lfloor\beta\rfloor+1}{\beta+1}$.
Putting together, we have the following remark.

\begin{pozn}
The numeration system with negative base $-\beta$ given by the transformation~\eqref{eq:cislo1} gives representation of numbers over the digit set
$\{0,1,\dots,\lfloor\beta\rfloor\}$ if and only if the left end-point of the domain $I=[l,r)$ of $T$ satisfies
\begin{equation}\label{eq:podminkacifry}
-\frac{\beta}{\beta+1}\leq l < -\frac{\lfloor\beta\rfloor+1}{\beta+1}\,.
\end{equation}
\end{pozn}

\begin{ex}\label{ex:itosadahiroproblem}
Putting $l=-\frac{\beta}{\beta+1}$ ensures that~\eqref{eq:podminkacifry} is satisfied universally for all $\beta$. Such choice of $l$
corresponds to the number system introduced in~\cite{ItoSadahiro} by Ito and Sadahiro. In this case, we have the transformation
\begin{equation}\label{eq:transformIS}
T:\Big[-\frac{\beta}{\beta+1},\frac{1}{\beta+1}\Big)\mapsto \Big[-\frac{\beta}{\beta+1},\frac{1}{\beta+1}\Big)\,,\qquad
T(x)=-\beta x - \Big\lfloor-\beta x + \frac{\beta}{\beta+1}\Big\rfloor\,.
\end{equation}
Note that the end-points of the interval $[l,r)$ now satisfy
$$
\frac{l}{-\beta} = l+1 = r\,.
$$
This fact is of certain help when deriving the condition on the admissibility of digit strings. On the other hand, it brings
one non-desirable phenomena. Just as in the usual positional number systems, we would expect that
multiplying by a power of the base only shifts the digit string representing the number. So
it would be natural to ask that
$$
\hbox{if }\quad \d(x)=x_1x_2x_3\cdots\,,\quad\hbox{ then }\quad \d\Big(\frac{x}{(-\beta)^k}\Big)=0^kx_1x_2x_3\cdots\,.
$$
This is however not true for all $x\in\big[-\frac{\beta}{\beta+1},\frac{1}{\beta+1}\big)$ in the Ito-Sadahiro
numeration system. Namely, taking for $x$ the boundary point $l=-\frac{\beta}{\beta+1}$ and denoting  $\d(l)=l_1l_2l_3\cdots$, we have
$(-\beta)^{-2}l\in[l,r)$ but
$$
\d\Big(\frac{l}{(-\beta)^2}\Big) = 1l_1l_2l_3\cdots\,,
$$
as can be verified easily using~\eqref{eq:transformIS}.
\end{ex}

The non-desirable phenomena occurring in the Ito-Sadahiro numeration system is caused by the fact that in the interval $[l,r)$ there
is an element, namely $l$ itself, which divided by $-\beta$ falls outside of $[l,r)$. Such phenomena can be avoided choosing such
$l$ that for given $\beta$ the following
implication holds
\begin{equation}\label{eq:cislo5}
x\in[l,r) \quad\implies\quad \frac{x}{-\beta}\in[l,r)\,.
\end{equation}
For, the transformation $T$ from~\eqref{eq:cislo1} satisfies $T(\frac{x}{-\beta})=x$ for all $x\in[l,r)$.
 Thus,
the implication is a necessary and sufficient condition so that the expansion of every real number $x$ is unique.
One verifies easily that implication~\eqref{eq:cislo5} is ensured by the following condition.

\begin{pozn}
The numeration system with negative base $-\beta$ given by the transformation~\eqref{eq:cislo1} gives a unique representation
$(-\beta,l)$-expansion of a non-zero $x\in\R$ in the form $x=\sum_{i=k}^\infty \frac{x_i}{(-\beta)^i}$, $x_k\neq 0$, if and only if
 the left end-point of the domain $I=[l,r)$ of $T$ satisfies
\begin{equation}\label{eq:shift}
-\frac{\beta}{\beta+1}<l \leq -\frac{1}{\beta+1}\,.
\end{equation}
\end{pozn}

\begin{ex}
A suitable choice of $l$ satisfying~\eqref{eq:shift} universally for all $\beta>1$, is $l=-\frac{1}{2}$. In this case, we obtain the digit set
$\Big\{ \big\lfloor\frac{-\beta+1}{2}\big\rfloor, \dots, \big\lfloor\frac{\beta+1}{2}\big\rfloor \Big\}$ which, for $\beta+1\notin 2\Z$, results
in the symmetric alphabet
$$
\Big\{ -\Big\lfloor\frac{\beta+1}{2} \Big\rfloor, \dots, \Big\lfloor\frac{\beta+1}{2}\Big\rfloor \Big\}\,.
$$
Consider the transformation $T(x)=-\beta x - \big\lfloor-\beta x + \frac{1}{2}\big\rfloor$. For all but finitely many $x$ from the interval
$[-\frac{1}{2},\frac{1}{2})$ we can write
\begin{equation}\label{eq:symetrie1/2}
\begin{aligned}
T(-x)&=-\beta (-x) - \Big\lfloor-\beta (-x) + \frac{1}{2}\Big\rfloor = - \Big( -\beta x -  \Big\lceil -\beta x - \frac{1}{2}\Big\rceil\Big) = \\
&=- \bigg( -\beta x - \Big(\Big\lceil -\beta x + \frac{1}{2}\Big\rceil-1\Big)\bigg) = - \Big( -\beta x - \Big\lfloor -\beta x + \frac{1}{2}\Big\rfloor\Big) = -T(x)\,.
\end{aligned}
\end{equation}
Here we have used that $\lfloor y\rfloor = - \lceil - y \rceil$ for all $y\in \R$ and $\lceil - y \rceil = \lfloor y\rfloor +1$ for all $y\notin \Z$. For $l=-\frac{1}{2}$, we have
$$
\d(-x) = \overline{\d(x)}\,,\quad \hbox{for all $x\in[-\frac{1}{2},\frac{1}{2})$ up to a countably many exceptions.}
$$
Let us specify that the notation $\overline{a}$ stands for the digit $-a$; similarly for a string of digits,
we write $\overline{a_1a_2\dots}$ meaning $\overline{a}_1\overline{a}_2\cdots = (-a_1)(-a_2)\cdots$.
Due to the latter property, we may call the numeration system with $l=-\frac12$ the balanced system.
\end{ex}

A very important practical requirement for the numeration system is that the set of numbers with finite expansions is non-trivial.
A number is said to have finite $(-\beta,l)$-expansion if $\d(x)$ ends in $0^\omega$. We denote the set of such
numbers by ${\rm Fin}(-\beta,l)$. As we have explained above, $0\in{\rm Fin}(-\beta,l)$ for any $\beta>1$ and any $l$ such that
$0\in[l,r)$. We can then write
$$
{\rm Fin}(-\beta,l) = \{x\in[l,r)\mid T^n(x)=0\text{ for some }n\in\N\}\,.
$$

However, for some choices of $l$ there are no other elements  with finite  $(-\beta,l)$-expansion besides zero.
The following proposition characterizes the numeration systems for which this does not happen.

\begin{prop}
The necessary and sufficient condition for ${\rm Fin}(-\beta,l)$ to be different from $\{0\}$ is that
\begin{equation}\label{eq:konecne}
-\frac{1}{\beta}\text{ or }\frac{1}{\beta}\in[l,r)\,.
\end{equation}
\end{prop}

\pfz
If a non-zero number belongs to ${\rm Fin}(-\beta,l)=\{x\in[l,r)\mid T^n(x)=0\text{ for some }n\in\N\}$, then necessarily, there exists
a non-zero number $y\in[l,r)$ such that $T(y)=0$. Since
$T(y)=-\beta y - \lfloor-\beta y -l\rfloor$,
the pre-images of 0 in the interval $[l,r)$ are of the form $\frac{a}{-\beta}$ for some integer $a$. On the other hand,
$T(\frac{a}{-\beta}) = a - \lfloor a-l \rfloor = 0$ for all $a$ such that $\frac{a}{-\beta}\in[r,l)$.
Obviously, $\frac{a}{-\beta}\in[r,l)$ for some non-zero integer $a$ if and only if $\frac{1}{\beta}$ or $-\frac{1}{\beta}$ belongs to $[l,r)$.
\pfk

Consider the Ito-Sadahiro numeration system. Here the condition~\eqref{eq:konecne} gives
${\rm Fin}\big(-\beta,-\frac{\beta}{\beta+1}\big)\neq\{0\}$ if and only if
$\beta\geq \tau$, where $\tau=\frac{1}{2}(1+\sqrt{5})$ is the golden ratio, as already mentioned in~\cite{MaPeVa}.
In the balanced numeration system we have ${\rm Fin}\big(-\beta,-\frac{1}{2}\big)\neq\{0\}$ if and only if $\beta\geq 2$.
On the other hand, we can choose a system which for any $\beta$ provides a non-trivial set of finite expansions. Such a system is obtained for example
if $l=-\frac{1}{\beta}$. Unfortunatelly, now the condition~\eqref{eq:shift} for uniqueness of the representation of numbers is not universally satisfied.
In particular, it is valid if and only if $\beta>\tau$.

%%%%%%%%%%%%%%%%%%%%%%%%%%%%%%%%%%%%%%%%%%%%%%%%%%%%%%%%%%%%%%%%%%%%%%%%%%%%%%%%%%%%%%%%%%%%%%%%%%%%%%%%%%%%%%%%%%%%%%%%%%
%\section{The $(-\beta,l)$-transformation}\label{sec:transformation}

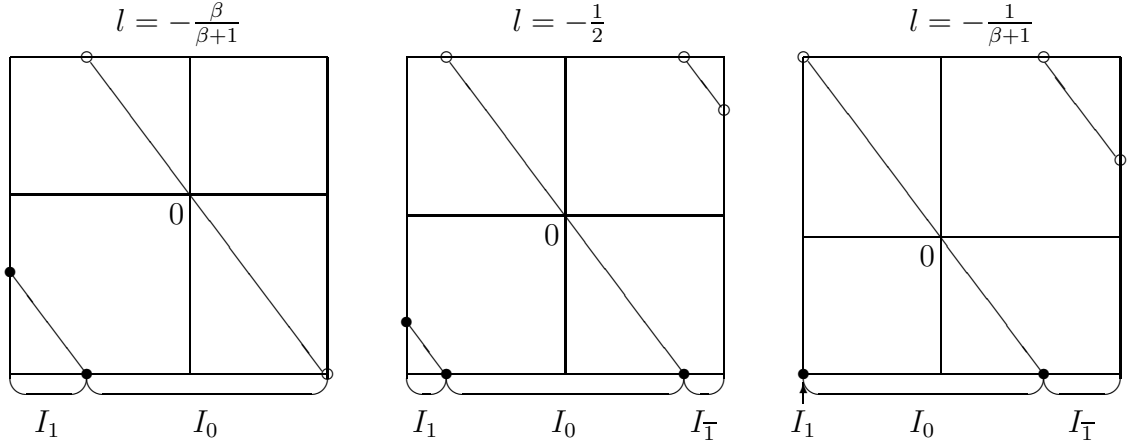
\begin{figure}[h]
\begin{center}
\begin{picture}(430,170)
%\put(0,0){\line(0,1){170}}\put(0,0){\line(1,0){430}}\put(430,170){\line(-1,0){430}}\put(430,170){\line(0,-1){170}}

%prvni graf minimal pisot IS
\put(5,25){\line(0,1){120}}\put(5,25){\line(1,0){120}}\put(125,145){\line(0,-1){120}}\put(125,145){\line(-1,0){120}} %ramecek
\put(73,25){\line(0,1){120}}\put(5,93){\line(1,0){120}} %osy
\put(35.5,143){\line(3,-4){87.5}}\put(34,25){\line(-3,4){28}} %transformace
\put(34,145){\circle{4}}\put(125,25){\circle{4}}
\put(34,25){\circle*{4}}\put(5,63.6){\circle*{4}} %puntiky
\put(65,82){$0$}
\put(14,3){$I_1$}\put(19.5,25){\oval(29,15)[b]}
\put(74,3){$I_0$}\put(79.5,25){\oval(91,15)[b]} %popisky
\put(45,155){$l=-\frac{\beta}{\beta+1}$}

%druhy graf minimal pisot balanced
\put(155,25){\line(0,1){120}}\put(155,25){\line(1,0){120}}\put(275,145){\line(0,-1){120}}\put(275,145){\line(-1,0){120}} %ramecek
\put(215,25){\line(0,1){120}}\put(155,85){\line(1,0){120}} %osy
\put(260,25){\line(-3,4){88.5}}\put(169,26){\line(-3,4){13}}\put(274,126){\line(-3,4){13}} %transformace
\put(170,145){\circle{4}}\put(260,145){\circle{4}}\put(275,125){\circle{4}}
\put(155,45){\circle*{4}}\put(170,25){\circle*{4}}\put(260,25){\circle*{4}} %puntiky
\put(207,74){$0$}
\put(262,3){$I_{\overline{1}}$}\put(267.5,25){\oval(15,15)[b]}
\put(209.5,3){$I_0$}\put(215,25){\oval(90,15)[b]}
\put(157,3){$I_1$}\put(162.5,25){\oval(15,15)[b]} %popisky
\put(195,155){$l=-\frac{1}{2}$}

%treti graf minimal pisot l=1/beta+1
\put(305,25){\line(0,1){120}}\put(305,25){\line(1,0){120}}\put(425,145){\line(0,-1){120}}\put(425,145){\line(-1,0){120}} %ramecek
\put(357,25){\line(0,1){120}}\put(305,77){\line(1,0){120}} %osy
\put(307,143.5){\line(3,-4){88.5}}\put(423,107.5){\line(-3,4){26.5}} %transformace
\put(305,145){\circle{4}}\put(396,145){\circle{4}}\put(425,106){\circle{4}}
\put(396,25){\circle*{4}}\put(305,25){\circle*{4}} %puntiky
\put(349,66){$0$}
\put(405,3){$I_{\overline{1}}$}\put(410.5,25){\oval(29,15)[b]}
\put(345,3){$I_0$}\put(350.5,25){\oval(91,15)[b]}
\put(300,3){$I_1$}\put(305,14){\vector(0,4){8}} %popisky
\put(345,155){$l=-\frac{1}{\beta+1}$}

\end{picture}
\caption{$(-\beta)$-transformations corresponding to $\beta$ minimal Pisot number (the real root of $x^3-x-1$) with different choices of domain $[l,r)$. For $l=-\frac{\beta}{\beta+1}$ it is the Ito-Sadahiro system, for $l=-\frac12$ it is the balanced system. In the third case, the interval $I_1=\{-\frac1{\beta+1}\}$ is degenerate, thus the corresponding digit 1 appears in $(-\beta,l)$-expansions of only countably many numbers.
In all the cases, the graph of the transformation intersects the line $y=0$ only at the origin, and thus the only element of ${\rm Fin(-\beta,l)}$
is 0.}
\label{f}
\end{center}
\end{figure}

\bigskip
Let us now study the properties of the $(-\beta,l)$-transformation $T$ of the interval $[l,r)$ in more depth. These properties will mainly be used in the proof of Theorem~\ref{thm:limitniretezce}. From the prescription
$T(x) = -\beta x - \lfloor - \beta x -l\rfloor$
using integer part we derive that, on its domain, $T$ is always continuous from the left, but there are discontinuity points in which
$T$ is discontinuous from the right, namely
\begin{equation}\label{eq:D}
{\mathcal D} = \bigg\{\frac{a+l}{-\beta} \ \bigg|\ a\in\A_{-\beta,l} \bigg\}\,.
\end{equation}
For illustration of the graph of $T$ see Figure~\ref{f}.
It is obvious that for a discontinuity point $x=\frac{a+l}{-\beta}\in{\mathcal D}$, we have
$$
T(x) = -\beta \frac{a+l}{-\beta} - \bigg\lfloor - \beta \frac{a+l}{-\beta} - l \bigg\rfloor = l\,.
$$
Moreover, we easily realize that
\begin{equation}\label{eq:discontinuity}
T(x) = l \quad\iff\quad x\in{\mathcal D}\,.
\end{equation}

The domain $[l,r)$ of the transformation $T$ is divided by the discontinuity points ${\mathcal D}$
into disjoint intervals $I_a$, $a\in\A_{-\beta,l}$, in such a way that for every $x\in I_a$, the first digit $x_1$ in the
$(-\beta,l)$-expansion of $x$ is equal to $a$.
Note that for the maximal digit $a=d_1$, the interval $I_{d_1}$ is closed, degenerate or non-degenerate. On the other hand, for the minimal
digit $a$, the interval $I_a$ is open. For the other digits, we always have the interval $I_a$ of the form $(\cdot,\cdot]$.

From the prescription~\eqref{eq:cislo1} we easily derive that fixed points of the $(-\beta,l)$-transformation are precisely the points
$f_a:=-\frac{a}{\beta+1}$, where $a$ is a digit in $\A_{-\beta,l}$. We thus have $\d(f_a)=a^\omega$ for every $a\in\A_{-\beta,l}$. Obviously, we have $f_a\in I_a$ and, moreover, $f_a$ is in most cases an inner point of $I_a$. In fact, it can lie on the boundary of $I_a$ only if
$$
\lfloor -\beta (f_a+\varepsilon) - l \rfloor < \lfloor -\beta f_a - l \rfloor = a \qquad\text{ for any $\varepsilon>0$}\,,
$$
and this happens only if $-\beta f_a - l = a$, i.e.\ $l=f_a$. In such a case, $a$ is the maximal digit and $I_a$ is a degenerate interval, and thus
the digit $a$ occurs in the $(-\beta,l)$-expansion of only countably many numbers in the domain $[l,r)$. It follows that always at least two among the intervals $I_a$, $a\in\A_{-\beta,l}$ are non-degenerate.

%From the prescription~\eqref{eq:cislo2}, we obtain for every digit $a\in\A_{-\beta,l}\subset\Z$ that
%\begin{equation}\label{eq:x}
%-(l+1)(\beta+1) < a \leq -l(\beta+1)\,.
%\end{equation}
%It follows directly that the points $f_a:=-\frac{a}{\beta+1}\in I_a$, $a\in\A_{-\beta,l}$, are fixed points of the transformation $T$, and consequently
%$\d(f_a)=a^\omega$ for every $a\in\A_{-\beta,l}$. If for a given digit $a$ in the inequality~\eqref{eq:x} both inequalities are strict,
%then $I_a$ is a non-degenerate interval and $f_a$ is its inner point.
%On the other hand, equality in~\eqref{eq:x} may happen only if $-l(\beta+1)$ is an integer. In that case
%$$
%l=-\frac{k}{\beta+1}\quad\text{ for some }\ k\in\{0,1,\dots,\lceil\beta\rceil\}
%$$
%and the interval $I_k=\{l\}$ is the only degenerate interval among $I_a$, $a\in\A_{-\beta,l}$. Moreover, the digit $k$ is then the maximal
%digit of the alphabet. Note that in such a case, the maximal digit $k$ occurs in the $(-\beta,l)$-expansion of only countably many
%numbers in the domain $[l,r)$.
%We will later use the obvious fact that always at least two among the intervals $I_a$, $a\in\A_{-\beta,l}$ are non-degenerate.

%\begin{pozn}\label{pozn:degenerovane_intervaly}
Consider $a\in\A_{-\beta,l}$ such that $I_a$ is non-degenerate. The corresponding fixed point $f_a$ of the transformation $T$ is an inner point of $I_a$, i.e.\ $T$ is continuous on some neighborhood of $f_a$. We can derive that for every $n\in\N$ there exists a positive $\delta_n$
satisfying $T^i(f_a-\delta_n,f_a+\delta_n)\subset I_a$ for $i=0,1,\dots,n-1$. Necessarily, for all $x\in(f_a-\delta_n,f_a+\delta_n)$, the
$(-\beta,l)$-expansion of $x$ starts with the prefix $a^n$. We have thus proved the following lemma which will be of use in Section~\ref{sec:periodic}.
%\end{pozn}

\begin{lem}\label{cl:pomocnyclaim}
For every $\beta>1$ and $l\in(-1,0]$ there exists a non-zero digit $a$ such that for all $n\in\N$ we
can find an interval $J\subset[l,r)$ such that the $(-\beta,l)$-expansion of every $x\in J$ is of the form
\begin{equation}\label{eq:pomocnyclaim}
\d(x) = a^nx_1x_2x_3\cdots\,.
\end{equation}
\end{lem}

%%%%%%%%%%%%%%%%%%%%%%%%%%%%%%%%%%%%%%%%%%%%%%%%%%%%%%%%%%%%%%%%%%%%%%%%%%%%%%%%%%%%%%%%%%%%%%%%%%%%%%%%%%%%
\section{Alternate order and admissible $\boldsymbol{(-\beta,l)}$-expansions}\label{sec:admis}

Let us show that, just as in the case of Ito-Sadahiro expansions, the alternate order on digit strings $\d(x)$ corresponds to the natural order
of real numbers in the interval $[l,r)$.

\begin{thm}\label{thm:altorder}
Alternate order on $(-\beta,l)$-expansions preserves the natural order on the interval $[l,r)$, i.e.
for $x,y\in[l,r)$ with $(-\beta,l)$-expansions $\d(x)$ and $\d(y)$ we have
$$
x<y \qquad\iff\qquad \d(x) \prec_{\hbox{\tiny alt}} \d(y)\,.
$$
\end{thm}

\pfz
For $x,y\in[l,r)$ with $(-\beta,l)$-expansions $\d(x)=x_1x_2x_3\cdots$ and $\d(y)=y_1y_2y_3\cdots$ we will show that
$$
x<y\quad\iff\quad x_m(-1)^m < y_m(-1)^m\,,
$$
where $m:=\min\{k\in\N\mid x_k\neq y_k\}$.
From the definition of $m$ and the construction of $(-\beta,l)$-expansions it follows that there exist $\tilde{x},\tilde{y}\in[l,r)$ and $z\in\R$
such that
$$
x=z+\frac{1}{(-\beta)^{m-1}}\tilde{x}\quad\hbox{and}\quad y=z+\frac{1}{(-\beta)^{m-1}}\tilde{y}\,,
$$
where $x_mx_{m+1}x_{m+2}\cdots$, $y_my_{m+1}y_{m+2}\cdots$ are the $(-\beta,l)$-expansions of $\tilde{x}$, $\tilde{y}$, respectively.
We thus have
\begin{equation}\label{eq:ekvival}
x<y\quad\iff\quad \frac{\tilde{x}}{(-\beta)^{m-1}} < \frac{\tilde{y}}{(-\beta)^{m-1}} \quad\iff\quad \tilde{x}(-1)^{m-1} < \tilde{y}(-1)^{m-1}\,.
\end{equation}
Since $x_m=\lfloor -\beta \tilde{x} - l \rfloor$, $y_m=\lfloor -\beta \tilde{y} - l \rfloor$, we have the following implications
$$
\begin{aligned}
x_m<y_m \quad\implies\quad  -\beta \tilde{x} - l <  -\beta \tilde{y} - l  \quad\implies\quad  \tilde{x} > \tilde{y}\,,\\
x_m>y_m \quad\implies\quad  -\beta \tilde{x} - l >  -\beta \tilde{y} - l  \quad\implies\quad  \tilde{x} < \tilde{y}\,,\\
\end{aligned}
$$
and thus
\begin{equation}\label{eq:ekvival2}
\tilde{x}>\tilde{y} \quad\iff\quad  x_m < y_m
\end{equation}
If $m$ is odd, equivalences~\eqref{eq:ekvival} and~\eqref{eq:ekvival2} imply
$x<y \Leftrightarrow \tilde{x}<\tilde{y} \Leftrightarrow  x_m > y_m \Leftrightarrow x_m(-1)^m < y_m(-1)^m$.
If $m$ is even, we have
$x<y \Leftrightarrow \tilde{x}>\tilde{y} \Leftrightarrow  x_m < y_m \Leftrightarrow x_m(-1)^m < y_m(-1)^m$.
%$$
%x<y \quad\iff\quad \tilde{x}<\tilde{y} \quad\iff\quad  x_m > y_m \quad\iff\quad x_m(-1)^m < y_m(-1)^m\,.
%$$
%If $m$ is even, we have
%$$
%x<y \quad\iff\quad \tilde{x}>\tilde{y} \quad\iff\quad  x_m < y_m \quad\iff\quad x_m(-1)^m < y_m(-1)^m\,.
%$$
\pfk

\begin{coro}\label{c:limity}
The $(-\beta,l)$-expansion as a function $\d:[l,r)\mapsto\Z^\N$ is strictly increasing on the interval $[l,r)$ in the alternate order,
and therefore the limits
$$
\d^*(l):=\lim_{\varepsilon\to0+}\d(l+\varepsilon)\,,\qquad
\d^*(r):=\lim_{\varepsilon\to0+}\d(r-\varepsilon)
$$
exist and $\d(x)\preceq_{\hbox{\tiny alt}}\d^*(l)$.
\end{coro}

The notation from Corollary~\ref{c:limity} will be used throughout the paper.

Theorem~\ref{thm:altorder} will now be used to prove a necessary and sufficient condition for admissibility of digit strings.

\begin{de}
An infinite string $x_1x_2x_3\cdots$ of integers is called $(-\beta,l)$-admissible (or just admissible), if there exists an $x\in[l,r)$ such that
$x_1x_2x_3\cdots$ is its $(-\beta,l)$-expansion, i.e. $x_1x_2x_3\cdots=\d(x)$.
\end{de}

%\begin{coro}\label{c:admis}
%Let $a_1a_2a_3\cdots$ be a $(-\beta,l)$-admissible digit string. Denote $\d(l)=l_1l_2l_3\cdots$ and
% $\d^*(r)=\lim_{\varepsilon\to0+}\d(r-\varepsilon)=r_1r_2r_3\cdots$. Then
%$$
%l_1l_2l_3\cdots\preceq_{\hbox{\tiny alt}} a_ia_{i+1}a_{i+2}\cdots  \prec_{\hbox{\tiny alt}} r_1r_2r_3\cdots\,,
%\qquad\hbox{for all $\ i\geq 1$}.
%$$
%\end{coro}
%
%In fact, the statement can be, just as in case of Ito-Sadahiro expansions, reversed.

\begin{thm}\label{thm:hlavni}
Let $d$ be the $(-\beta,l)$-expansion and denote $\d(l)=l_1l_2l_3\cdots$ and $\d^*(r)=r_1r_2r_3\cdots$.
An infinite string $x_1x_2x_3\cdots$ of integers is $(-\beta,l)$-admissible, if and only if
\begin{equation}\label{eq:admis}
l_1l_2l_3\cdots\preceq_{\hbox{\tiny alt}} x_ix_{i+1}x_{i+2}\cdots  \prec_{\hbox{\tiny alt}} r_1r_2r_3\cdots\,,
\qquad\hbox{for all $\ i\geq 1$}.
\end{equation}
\end{thm}

\pfz
The fact that the condition~\eqref{eq:admis} is necessary follows as a consequence of Theorem~\ref{thm:altorder}.
It remains to show that it is sufficient, i.e. that a digit string $x_1x_2x_3\cdots$ satisfying
the condition~\eqref{eq:admis} is a $(-\beta,l)$-expansion of an $x\in[l,r)$.
It is sufficient to show that for all $n=1,2,3,\dots$ we have
\begin{equation}\label{eq:nerovnost}
{\scriptstyle \bullet}x_nx_{n+1}x_{n+2}\cdots = \sum_{i=1}^\infty \frac{x_{n-1+i}}{(-\beta)^i}\in[l,r)\,.
\end{equation}
where we use the notation ${\scriptstyle \bullet}y_1y_{2}y_{3}\cdots$ for the number represented
by the digit string $y_1y_2y_3\cdots$, i.e. for the value
$$
{\scriptstyle \bullet}y_1y_{2}y_{3}\cdots = \frac{y_1}{(-\beta)} + \frac{y_2}{(-\beta)^2}+\frac{y_3}{(-\beta)^3}+\cdots
= \sum_{i=1}^\infty \frac{y_{i}}{(-\beta)^i}\,.
$$
Let us fix an $n\in\N$. Since $x_nx_{n+1}x_{n+2}\cdots \prec_{\hbox{\tiny alt}} r_1r_2r_3\cdots=\d^*(r)$, there
exists an $\varepsilon>0$ such that
$x_nx_{n+1}x_{n+2}\cdots \prec_{\hbox{\tiny alt}} \d(r-\varepsilon)$. Choose an $\varepsilon>0$  such that
also
\begin{equation}\label{eq:8}
x_{n+1}x_{n+2}x_{n+3}\cdots \prec_{\hbox{\tiny alt}} \d(r-\varepsilon)=\tilde{r}_1\tilde{r}_2\tilde{r}_3\cdots\,.
\end{equation}
We first show by induction on $s$ the following statement:
%$$
%(\forall s\in\N)(\forall z\in I \hbox{ with $(-\beta,l)$-expansion $z_1z_2z_3\cdots$ })(\forall n\in\N) \hbox{ we have}
%$$

\begin{claim}\label{claim}
Let $z\in [l,r)$ have the $(-\beta,l)$-expansion of the form $\d(z)=z_1z_2z_3\cdots$ and let $s,m\in\N$.

\noindent
If \ $x_nx_{n+1}\cdots x_{n+s}0^\omega\succeq_{\hbox{\tiny alt}} z_mz_{m+1}\cdots z_{m+s}0^\omega$, \
then \ ${\scriptstyle  \bullet}x_nx_{n+1}\cdots x_{n+s} \geq {\scriptstyle  \bullet} z_mz_{m+1}\cdots  - \frac{1}{\beta^{s+1}}$.

\noindent
If \ $x_nx_{n+1}\cdots x_{n+s}0^\omega\preceq_{\hbox{\tiny alt}} z_mz_{m+1}\cdots z_{m+s}0^\omega$, \ then \
${\scriptstyle  \bullet}x_nx_{n+1}\cdots x_{n+s} \leq {\scriptstyle  \bullet} z_mz_{m+1}\cdots  + \frac{1}{\beta^{s+1}}$.
\end{claim}

\pfz
For $s=0$, the inequality $x_n0^\omega\succeq_{\hbox{\tiny alt}} z_m0^\omega$ is equivalent to $x_n\leq z_m$. We thus have
$$
\frac{x_n}{-\beta}\geq\frac{z_m}{-\beta} = {\scriptstyle  \bullet} z_mz_{m+1}\cdots +
\frac{1}{\beta}\underbrace{({\scriptstyle  \bullet} z_{m+1}z_{m+2}\cdots)}_{=T^{m+1}(z)\in(-1,1)} \geq
{\scriptstyle  \bullet} z_mz_{m+1}\cdots - \frac{1}{\beta}\,.
$$
For $s\geq 1$, the inequality $x_nx_{n+1}\cdots x_{n+s}0^\omega\succeq_{\hbox{\tiny alt}} z_mz_{m+1}\cdots z_{m+s}0^\omega$ implies that one of the following happens,
\begin{itemize}
\item[(i)] either $x_n=z_m$ and $x_{n+1}\cdots x_{n+s}0^\omega\preceq_{\hbox{\tiny alt}} z_{m+1}\cdots z_{m+s}0^\omega$,
\item[(ii)] or $x_n\leq z_m-1$.
\end{itemize}

In case (i) that $x_n=z_m$, we use induction hypothesis
to derive from $x_{n+1}\cdots x_{n+s}0^\omega\preceq_{\hbox{\tiny alt}} z_{m+1}\cdots z_{m+s}0^\omega$ that
$$
{\scriptstyle  \bullet}x_{n+1}\cdots x_{n+s} \leq {\scriptstyle  \bullet} z_{m+1}\cdots  + \frac{1}{\beta^{s}}\,.
$$
Dividing by $-\beta$ and adding $x_n(-\beta)^{-1}=z_m(-\beta)^{-1}$, we have
$$
{\scriptstyle  \bullet}x_nx_{n+1}\cdots x_{n+s} \geq {\scriptstyle  \bullet} z_mz_{m+1}\cdots  - \frac{1}{\beta^{s+1}}\,.
$$

In case (ii), we have $z_m-x_n\geq 1$ and thus
\begin{equation}\label{eq:pomoc1}
\frac{z_m-x_n}{-\beta}\leq -\frac{1}{\beta}\,.
\end{equation}
For the string $x_{n+1}\cdots x_{n+s}$ we can derive using~\eqref{eq:8} that for prefixes $l_1l_2\cdots l_s$ of
$\d(l)$ and $\tilde{r}_1\tilde{r}_2\cdots \tilde{r}_s$ of $\d(r-\varepsilon)$ it holds that
$$
l_1l_2\cdots l_s0^\omega \preceq_{\hbox{\tiny alt}} x_{n+1}\cdots x_{n+s}0^\omega \preceq_{\hbox{\tiny alt}} \tilde{r}_1\tilde{r}_2\cdots \tilde{r}_s0^\omega\,.
$$
By induction hypothesis, we have
\begin{equation}\label{eq:pomoc2}
l-\frac{1}{\beta^s} \leq {\scriptstyle  \bullet}x_{n+1}\cdots x_{n+s} \leq r-\varepsilon + \frac{1}{\beta^s}
\end{equation}
We therefore have
$$
\begin{aligned}
{\scriptstyle  \bullet}x_nx_{n+1}\cdots x_{n+s} - {\scriptstyle  \bullet} z_mz_{m+1}\cdots  &=
\underbrace{\frac{x_n-z_m}{-\beta}}_{\geq \frac{1}{\beta}} -\frac{1}{\beta}\Big(\underbrace{{\scriptstyle  \bullet}x_{n+1}\cdots x_{n+s}}_{\leq r-\varepsilon+\frac{1}{\beta^s}} - \underbrace{{\scriptstyle  \bullet} z_{m+1}z_{m+2}\cdots}_{\geq l}\Big) \geq\\
&\geq \frac{1}{\beta} - \frac{1}{\beta}\Big(r-\varepsilon+\frac{1}{\beta^{s}}- l\Big) = -\frac{1}{\beta^{s+1}} + \frac{\varepsilon}{\beta}\geq
-\frac{1}{\beta^{s+1}}\,,
\end{aligned}
$$
where we have used~\eqref{eq:pomoc1} and~\eqref{eq:pomoc2}. By this, the claim is proved.
\pfk

Let us now use Claim~\ref{claim} for proving the statement of Theorem~\ref{thm:hlavni}. For $z\in [l,r)$ of the claim we put $z=l$ and $z= r-\varepsilon$.
For all $s\in\N$, we have
$$
l_1l_2\cdots l_{s}0^\omega \preceq_{\hbox{\tiny alt}} x_nx_{n+1}\cdots x_{n+s}0^\omega \preceq_{\hbox{\tiny alt}} \tilde{r}_1\tilde{r}_2\cdots \tilde{r}_s0^\omega\,,
$$
and therefore by claim,
$$
l-\frac{1}{\beta^s} \leq {\scriptstyle  \bullet}x_nx_{n+1}\cdots x_{n+s-1} \leq r-\varepsilon + \frac{1}{\beta^s}\,.
$$
As $s$ tends to infinity, we derive
$$
l\leq {\scriptstyle  \bullet}x_nx_{n+1}x_{n+2}\cdots \leq r-\varepsilon < r\,,
$$
which is equivalent to~\eqref{eq:nerovnost} that we were to show.
\pfk

%\begin{pozn}\label{pozn:IS}
%Ito and Sadahiro have described the admissibility condition for the numeration system considered by them, i.e.\ for $l=-\frac{\beta}{\beta+1}$.
%They have shown that in their case the reference strings used in the condition in Theorem~\ref{thm:hlavni} (i.e. $\d(l)=l_1l_2l_3\cdots$ and $\d^*(r)=r_1r_2r_3\cdots$) are related in the following way:
%$$
%r_1r_2r_3\cdots = 0l_1l_2l_3\cdots
%$$
%if $\d(l)$ is not purely periodic with odd period length, and,
%$$
%r_1r_2r_3\cdots = \big(0l_1l_2\cdots l_{q-1}(l_{q}-1)\big)^\omega\,,
%$$
%if $\d(l) = \big(l_1l_2\cdots l_{q}\big)^\omega$, where $q$ is odd.
%\end{pozn}
%
%In what follows we describe
%some properties of the digit sequences occurring in the admissibility condition
%as left and right reference strings. We also derive the exact prescription for their relation in some cases.
%The Ito-Sadahiro instance is one of them.

%%%%%%%%%%%%%%%%%%%%%%%%%%%%%%%%%%%%%%%%%%%%%%%%%%%%%%%%%%%%%%%%%%%%%%%%%%%%%%%%%%%%%%%%%%%%%%%%%%%%%%%%%%%%%%%%%%%%%
%\section{Admissibility condition depending on the domain of the transformation $\boldsymbol{T}$}\label{sec:admisKonkret}
\section{Reference strings $\d(l)$ and $\d^*(r)$}\label{sec:digitstrings}

Let us now study the digit strings which appear in the $(-\beta,l)$-admissibility criteria. The following theorem states the relation between
$\d(l)$ and the limit $\d^*(l)=\lim\limits_{\varepsilon\to0+}\d(l+\varepsilon)$. Although $\d^*(l)$ itself does not appear in the admissibility condition,
it may, as we shall see on examples below, be used to derive the form of $\d^*(r)$.

\begin{thm}\label{thm:limitniretezce}
Let $T$ be the $(-\beta,l)$-transformation. If $T^q(l)\neq l$ for all $q\in\N$, or the equality $T^q(l)=l$ occurs only
for even $q\in\N$, then
$$
\d^*(l) = \d(l)\,.
$$
If on the other hand $T^q(l)= l$ for some $q\in\N$, $q$ odd, i.e.\
$\d(l)=(l_1l_2\cdots l_{q-1}l_q)^\omega$, then
$$
\d^*(l) = l_1l_2\cdots l_{q-1}(l_q-1)\d^*(r)\,.
$$
\end{thm}

\pfz
The proof uses the properties of the $(-\beta,l)$-transformation stated in Section~\ref{sec:transformation}.
Recall that the $(-\beta,l)$-transformation is continuous on $[l,r)$ except in the discontinuity points forming the set ${\mathcal D}$
given by~\eqref{eq:D}. The discontinuity points divide the interval $[l,r)$ into disjoint union
$[l,r)=\bigcup\limits_{a\in\A_{-\beta,l}} I_a$, where $I_a=\big\{x\in[l,r)\mid \text{$a$ is a prefix of $\d(x)$}\big\}$.
By~\eqref{eq:discontinuity} we have that $T(x)=l$ if and only if $x\in{\mathcal D}$.
Suppose that for all $i\leq k$ one has $T^i(l)\neq l$. Then $T^{k-1}(l)\notin {\mathcal D}$. Denote by $M$ the set
$$
M = \{T^{i}(l) \mid i=0,1,\dots, k-1\}\,.
$$
Choose $\delta>0$ such that
\begin{equation}\label{eq:kq}
\beta^k\delta < \min\{|x-y| \mid x\in M,\ y\in{\mathcal D}\}.
\end{equation}
For such $\delta$, we have $[l,l+\delta) \subset I_{l_1}$,
$$
T\big([l,l+\delta)\big) = \big(T(l) -\beta\delta, T(l) \big]\subset I_{l_{2}}\,,
$$
and more generally, we have for $2i-1\leq k$ that
\begin{equation}\label{eq:iteraceodd}
T^{2i-1}\big([l,l+\delta)\big) = \big(T^{2i-1}(l) -\beta^{2i-1}\delta, T^{2i-1}(l) \big]\subset I_{l_{2i}}\,,
\end{equation}
and for $2i\leq k$ that
\begin{equation}\label{eq:iteraceeven}
T^{2i}\big([l,l+\delta)\big) = \big[T^{2i}(l), T^{2i}(l)+\beta^{2i}\delta \big)\subset I_{l_{2i+1}}\,.
\end{equation}
Since
$T^j\big([l,l+\delta)\big) \subset I_{l_{j+1}}$ for every $j\leq k$, the $(-\beta,l)$-expansion $\d(x)$ of every $x\in[l,l+\delta)$
has the prefix $l_1l_2\cdots l_{k+1}$.

Suppose now that $T^i(l)\neq l$ for any $i\in\N$. Then obviously
$$
\d^*(l)= \lim_{\varepsilon\to0+} \d(l+\varepsilon) = \d(l)\,.
$$

Suppose on the other hand that one can find $q$ such that $T^q(l)=l$. Let $q$ be the minimal exponent with this property. We can
find $\delta$ so that~\eqref{eq:kq} is satisfied with $k=q-1$.
One derives that $T^{q-1}\big([l,l+\delta)\big)\subset I_{l_q}$ and $T^{q-1}(l)\in{\mathcal D}$.

First let $q-1$ be odd. By~\eqref{eq:iteraceodd} we have
$$
T^{q-1}\big([l,l+\delta)\big) = \big(T^{q-1}(l) -\beta^{q-1}\delta, T^{q-1}(l) \big]\subset I_{l_{q}}\,.
$$
Although $T^{q-1}(l)\in{\mathcal D}$, the transformation $T$ is continuous from the left at any discontinuity point, and so
$$
T^{q}\big([l,l+\delta)\big) = \big[T^{q}(l),T^q(l) +\beta^{q}\delta \big) = [l,l+\beta^q\delta)\,.
$$
This implies that for an arbitrarily long prefix $w$ of $\d(l)$ one finds an $\varepsilon>0$ such that for all $x\in[l,l+\varepsilon)$,
the $(-\beta,l)$-expansion of $x$ has the prefix $w$. In other words,
$$
\d^*(l) = \lim_{\varepsilon\to0+} \d(l+\varepsilon) = \d(l)\,.
$$

Let now $q-1$ be even. By~\eqref{eq:iteraceeven} we have
$$
T^{q-1}\big([l,l+\delta)\big) = \big[T^{q-1}(l), T^{q-1}(l)+\beta^{q-1}\delta \big)\subset I_{l_{q}}\,.
$$
Now $T^{q-1}(l)\in{\mathcal D}$ and $T$ is not continuous from the right, and thus
$$
T^{q}\big([l,l+\delta)\big) = \{l\} \cup (r-\beta^q\delta , r)\,.
$$
This implies that for all $0<\varepsilon<\delta$, the $(-\beta,l)$-expansion of $x=l+\varepsilon$ is of the form
$$
\d(x) = l_1l_2\cdots l_{q-1}(l_q-1)x_{q+1}x_{q+2}\cdots\,,
$$
where
$$
x_{q+1}x_{q+2}\cdots = \d(r-\beta^q\varepsilon)\,.
$$
Therefore
$$
\d^*(l)=\lim_{\varepsilon\to0+}\d(l+\varepsilon) = l_1l_2\cdots l_{q-1}(l_q-1) \d^*(r)\,.
$$
\pfk

We now illustrate the use of Theorem~\ref{thm:limitniretezce} for deriving the exact prescription for the relation
of the reference strings $\d(l)$ and $\d^*(r)$ for three choices of $l$.
First, let us take the example of the Ito-Sadahiro numeration system.

\begin{ex}\label{ex:mezeIS}
Let $l=-\frac{\beta}{\beta+1}$. We can compute
$$
T(r-\varepsilon) = -\beta\Big(-\frac{\beta}{\beta+1} + 1\Big) + \beta\varepsilon - \underbrace{\Big\lfloor\frac{\beta^2}{\beta+1}-\beta+\beta\varepsilon + \frac{\beta}{\beta+1}\Big\rfloor}_{=\lfloor\beta\varepsilon\rfloor=0} = -\frac{\beta}{\beta+1} + \tilde{\varepsilon} = l+\tilde{\varepsilon}\,.
$$
Therefore
\begin{equation}\label{eq:ISpravamez}
\d^*(r)=0\d^*(l)\,.
\end{equation}
If, moreover, $\d(l)=(l_1\cdots l_q)^\omega$ is purely periodic with odd period length $q$, then using~\eqref{eq:ISpravamez} and Theorem~\ref{thm:limitniretezce}
we have
$$
\d^*(r)=0l_1l_2\cdots l_{q-1}(l_q-1)\d^*(r)\,,
$$
which implies
$$
\d^*(r)=\big(0l_1l_2\cdots l_{q-1}(l_q-1)\big)^\omega\,.
$$
If, on the other hand, $\d(l)$ is not purely periodic with odd period,
then~\eqref{eq:ISpravamez} implies that simply
$\d^*(r)=0\d(l)$. This together
corresponds to the result of Ito and Sadahiro cited in the introduction.
\end{ex}

\begin{ex}
Let $l=-\frac{1}{2}$. As we have shown in~\eqref{eq:symetrie1/2}, we have $T(-x)=-T(x)$ for almost all $x\in [l,r)$.
It follows that
$$
\overline{\d^*(l)} = \d^*(r)\,.
$$
In case that $\d(l)$ is not purely periodic with odd period length, then
$$
\d^*(r) = \overline{\d^*(l)} = \overline{\d(l)}\,.
$$
On the other hand, if $\d(l)=(l_1\cdots l_q)^\omega$ is purely periodic with odd period length $q$, then using Theorem~\ref{thm:limitniretezce}
we obtain
$$
\d^*(r) = \overline{l_1\cdots l_{q-1}(l_q-1)\d^*(r)}\,,
$$
and by iteration
$$
\d^*(r) = \overline{l_1\cdots l_{q-1}(l_q-1)}l_1\cdots l_{q-1}(l_q-1)\d^*(r)\,,
$$
which implies
$$
\d^*(r) = \Big(\overline{l_1\cdots l_{q-1}(l_q-1)}l_1\cdots l_{q-1}(l_q-1)\Big)^\omega
$$
\end{ex}

\begin{ex}
Let $l=-\frac{1}{\beta}$. In that case $l_1=\big\lfloor -\beta (-\frac{1}{\beta}) + \frac{1}{\beta} \big\rfloor = 1$ and
$T(l)=-\beta\big(-\frac{1}{\beta}\big) - 1 = 0$. Consequently, the $(-\beta,-\frac{1}{\beta})$-expansion of $l$ is $\d(l)=10^\omega$ for every base $-\beta$.
Since it is not purely periodic, we have for the left limit that $\d^*(l) = \d(l)$.
In this case, the right limit $\d^*(r)$ has no relation to $\d(l)$.
\end{ex}

%%%%%%%%%%%%%%%%%%%%%%%%%%%%%%%%%%%%%%%%%%%%%%%%%%%%%%%%%%%%%%%%%%%%%%%%%%%%%%%%%%%%%%%%%%%%%%%%%%%%%%%%%%%%%%%%%%%%%%%%%%%
%\section{Reference strings $\d(l)$ and $\d^*(r)$}\label{sec:digitstrings}

In the remaining part of this section we focus on the question what sequences may play role of the left and right reference strings in the admissibility condition.
We describe some properties of these digit sequences, but, as we shall see, the question about a necessary a sufficient condition is far from being solved even for the case of Ito-Sadahiro numeration system. We explain the phenomenon on two examples which, in fact, represent an impeachment to Theorem~25 of G\'ora~\cite{gora} who approaches this problem in a more general setting.

We start by a simple consequence of Theorem~\ref{thm:hlavni}.

\begin{coro}\label{c:admiskraje}
Let $d$ be the $(-\beta,l)$-expansion.
Every suffix of $\d(l)=l_1l_2l_3\cdots$ satisfies the admissibility condition~\eqref{eq:admis}, i.e.
\begin{equation}\label{eq:admisLevyKraj}
l_1l_2l_3\cdots\preceq_{\hbox{\tiny alt}} l_i l_{i+1} l_{i+2}\cdots  \prec_{\hbox{\tiny alt}}r_1r_2r_3\cdots\,,
\qquad\hbox{for all $\ i\geq 1$}.
\end{equation}
Every suffix of $\d^*(r)=\lim\limits_{\varepsilon\to0+}\d(r-\varepsilon)=r_1r_2r_3\cdots$ satisfies
\begin{equation}\label{eq:admisPravyKraj}
l_1l_2l_3\cdots\preceq_{\hbox{\tiny alt}} r_i r_{i+1} r_{i+2}\cdots  \preceq_{\hbox{\tiny alt}}r_1r_2r_3\cdots\,,
\qquad\hbox{for all $\ i\geq 1$}.
\end{equation}
\end{coro}

\pfz
The admissibility condition for $\d(l)$ is an obvious consequence of Theorem~\ref{thm:hlavni}.

In order to show the inequalities for $\d^*(r)$, realize that every suffix $r_i r_{i+1}r_{i+2}\cdots$
of $\d^*(r)$ is a limit of $(-\beta,l)$-expansions of numbers $T^{i-1}(r-\varepsilon)$, as $\varepsilon$ tends to 0 from the right. Since by Theorem~\ref{thm:hlavni} we have
$$
\d\big(T^{i-1}(r-\varepsilon)\big) \prec_{\hbox{\tiny alt}} \d^*(r)\,.
$$
Applying the limit, the strict inequality $\prec_{\hbox{\tiny alt}}$ becomes $\preceq_{\hbox{\tiny alt}}$.
%
%In order to show the inequalities for $\big(d_{-\beta}^c\big)^*(c+1)$, realize that
%$\big(d_{-\beta}^c\big)^*(c+1)$ is a limit of $(-\beta,c)$-expansions of numbers $c+1-\varepsilon$, as $\varepsilon$ tends to 0 from the right.
%Therefore, for every $n\in\N$ one finds an $\varepsilon_n>0$ such that
%$d_{-\beta}^c(c+1-\varepsilon_n)$ and $\big(d_{-\beta}^c\big)^*(c+1)$ have a common prefix of length at least $n$.
%If~\eqref{eq:admisPravyKraj} is not satisfied for some $i\geq 1$, then it is not satisfied for some prefix of
%$\tilde{d}_i\tilde{d}_{i+1}\tilde{d}_{i+2}\cdots \tilde{d}_{n}$. However, the inequalities are satisfied by $d_{-\beta}^c(c+1-\varepsilon_n)$,
%which leads to a contradiction.
\pfk

The admissibility condition of Corollary~\ref{c:admiskraje} has interesting implications for repetition of blocks in prefixes of the reference digit strings
$\d(l)$ and $\d^*(r)$.

\begin{coro}
If $w$ is a block of digits in the alphabet $\A$ of odd length such that $ww$ is a prefix of $\d(l)$, then $\d(l)=w^\omega$.

If $w$ is a block of digits in the alphabet $\A$ of odd length such that $ww$ is a prefix of $\d^*(r)$, then $\d^*(r)=w^\omega$.
\end{coro}

\pfz
Let us write
\begin{equation}\label{eq:dukazlevykraj}
\d(l)= \underbrace{ww\cdots w}_{\text{$k$ times}}vs\,,
\end{equation}
where $v$ is a string over $\A$ of the same length as $w$, $k\geq 2$ and $s$ is a suffix of $\d(l)$.
Admissibility condition in Corollary~\ref{c:admiskraje} implies
$$
\underbrace{ww\cdots w}_{\text{$k$ times}}vs \preceq_{\hbox{\tiny alt}} wvs
$$
Now we use the simple fact that the alternate order satisfies
$$
x \preceq_{\hbox{\tiny alt}} y \quad\iff\quad wx\succeq_{\hbox{\tiny alt}}wy
$$
for a string $w$ of odd length and any strings $x,y$. For our case this implies
$$
\underbrace{ww\cdots w}_{\text{$k-1$ times}}vs \succeq_{\hbox{\tiny alt}} vs
$$
and thus $w0^\omega\succeq_{\hbox{\tiny alt}}v0^\omega$. On the other hand, $w$ is the smallest among
prefixes of the same length of digit strings $\d(x)$ for all $x\in[l,r)$, and therefore
$w0^\omega\preceq_{\hbox{\tiny alt}} v0^\omega$. This implies that $w=v$ and we may deduce
that $\d(l)$  has prefix $w^{k+1}$. The procedure may be repeated and hence
$\d(l)=w^\omega$.

The same argument is used for the right reference string $\d^*(r)$.
\pfk

\begin{pozn}
As a result, we can observe that whenever $\d(l)$ starts with $aa$ for some digit $a$, then necessarily
$\d(l)=a^\omega$ and consequently $l=-\frac{a}{\beta+1}$.
%which is the case mentioned in Remark~\ref{pozn:degenerovane_intervaly}.
Similarly, prefix $bb$ in the right reference string implies that $\d^*(r)=b^\omega$.
%This, however, would mean that $c+1-\varepsilon$ has expansion of the form
%$d_{-\beta}^c(c+1-\varepsilon)=\tilde{d}^nx_{n+1}x_{n+2}\cdots$ where $x_{n+1}>\tilde{d}$, since $\tilde{d}$
%is the minimal digit in the admissible alphabet. Such string, would not be admissible, since one of the inequalities
%$$
%\begin{aligned}
%\tilde{d}^nx_{n+1}x_{n+2}\cdots &\prec_{\hbox{\tiny alt}} \tilde{d}^\omega\\
%\tilde{d}^{n-1}x_{n+1}x_{n+2}\cdots &\prec_{\hbox{\tiny alt}} \tilde{d}^\omega
%\end{aligned}
%$$
%fails, dependently on the parity of $n$. Therefore
%$\big(d_{-\beta}^c\big)^*(c+1) = \tilde{d}_1\tilde{d_2}\tilde{d}_3\cdots$ satisfies $\tilde{d_2}>\tilde{d_1}$.
\end{pozn}

Corollary~\ref{c:admiskraje} above shows that the bounds used in the admissibility condition satisfy the condition by themselves, i.e.\
the admissibility conditions~\eqref{eq:admisLevyKraj} and~\eqref{eq:admisPravyKraj} are a necessary condition so that strings
$l_1l_2l_3\cdots$ and $r_1r_2r_3\cdots$ coincide with the reference strings $\d(l)$ and $\d^*(r)$ corresponding to some $\beta>1$
and $l\in(-1,0]$. This is analogous to the fact that in the case of positive base $\beta$ the R\'enyi expansion of $1$ satisfied the Parry condition.
Let us recall that for base $\beta>1$ one uses the transformation $T(x)=\beta x-\lfloor \beta x\rfloor$ of the interval $[0,1)$. The string of integers
$d_\beta^*(1)=t_1t_2t_3\cdots$ must satisfy
$$
0^\omega \preceq_{\hbox{\tiny lex}} t_it_{i+1}t_{i+2}\cdots \preceq_{\hbox{\tiny lex}} t_1t_2t_3\cdots \quad\text{for every }i=1,2,3,\dots\,.
$$
Parry~\cite{Parry} has shown that this condition is also sufficient, in order that a digit string $t_1t_2t_3\cdots$ is equal to $d_\beta^*(1)$ for
some $\beta>1$.

A natural question arises, whether satisfying conditions
\begin{equation}\label{eq:gora}
\begin{array}{lclcl}
l_1l_2l_3\cdots &\preceq_{\hbox{\tiny alt}}& l_il_{i+1}l_{i+2}\cdots &\prec_{\hbox{\tiny alt}} & r_1r_2r_3\cdots\\
l_1l_2l_3\cdots &\preceq_{\hbox{\tiny alt}}& r_ir_{i+1}r_{i+2}\cdots &\preceq_{\hbox{\tiny alt}} & r_1r_2r_3\cdots
\end{array}\quad\text{for every }i=1,2,3,\dots\,,
\end{equation}
for a pair of integer sequences $(l_i)_{i=1}^\infty$, $(r_i)_{i=1}^\infty$ already ensures the existence of $l$ and $\beta$,
so that the sequences are equal to the left and right reference strings $\d(l)$ and $\d^*(r)$ corresponding to some $\beta>1$ and $l\in(-1,0]$.

Such question was addressed by G\'ora. In~\cite{gora}, he studies a more general number systems. However, a special case of his is equivalent to
representation with negative base. Theorem~25 in~\cite{gora} can be transformed into the statement that conditions~\eqref{eq:gora}
are sufficient for $(l_i)_{i=1}^\infty$, $(r_i)_{i=1}^\infty$ being the left and right reference strings for some $\beta$ and $l$.
Here we show that his statement cannot be true.

%However, in the positive base case, the R\'enyi expansion of $1$ was characterized
%by the lexicographical requirement that every suffix of it is lexicographically strictly smaller than the sequence itself.
%As we illustrate here, similar statement is not possible in the case of negative base.

In order to clarify the problem, let us focus on the Ito-Sadahiro numeration system, where $l=-\frac{\beta}{\beta+1}$ and the sequences
$\d(l)$ and $\d^*(r)$ are connected as recalled in Example~\ref{ex:mezeIS}, i.e.\ when $\d(l)$ is non-periodic, then $\d^*(r) = 0\d(l)$.

\begin{ex}
Consider the sequence $(x_i)_{i=1}^\infty=a(a-1)0a^\omega$ for some positive integer $a>1$. It is easily verified that every suffix $x_ix_{i+1}x_{x+2}\cdots$ of this digit string satisfies
\begin{equation}\label{eq:ispodminka}
a(a-1)0a^\omega \preceq_{\hbox{\tiny alt}} x_ix_{i+1}x_{i+2}\cdots \prec_{\hbox{\tiny alt}} 0 a(a-1)0 a^\omega\,.
\end{equation}
Now if it holds that $a(a-1)0a^\omega$ is the left reference string of the Ito-Sadahiro system, then
$$
-\frac{\beta}{\beta+1} = \frac{a}{-\beta} + \frac{a-1}{(-\beta)^2} + \frac{a}{(-\beta)^4} + \frac{a}{(-\beta)^5} + \cdots\,,
$$
which implies that $\beta=a$. However, applying the Ito-Sadahiro transformation $T$ with the domain $\big[-\frac{\beta}{\beta+1},\frac{1}{\beta+1}\big)$  to the point $-\frac{\beta}{\beta+1} = -\frac{a}{a+1}$ yields
$\d(-\frac{\beta}{\beta+1}) = a^\omega$. Thus, $a(a-1)0a^\omega$ is not an admissible left reference string,
although it satisfies~\eqref{eq:ispodminka}.
\end{ex}

The following example shows that such a case is not isolated and need not correspond to integer base $-\beta$.

\begin{ex}\label{ex:wolfik}
Consider in the Ito-Sadahiro system the digit string $200(21)^\omega$.  One easily verifies that
every suffix $x_ix_{i+1}x_{i+2}\cdots$ of $200(21)^\omega$ satisfies
$$
200(21)^\omega \preceq_{\hbox{\tiny alt}} x_ix_{i+1}x_{i+2}\cdots \prec_{\hbox{\tiny alt}} 0200(21)^\omega\,.
$$
Comparing
$$
-\frac{\beta}{\beta+1} = \frac{2}{-\beta} + \frac{2}{(-\beta)^4} + \frac{1}{(-\beta)^5} + \frac{2}{(-\beta)^6} +  \frac{1}{(-\beta)^7} +\cdots\,,
$$
leads to $\beta = \frac12(3+\sqrt5)$. However, for such a base, the left reference string in the Ito-Sadahiro number system is equal to
$(21)^\omega$.
\end{ex}

%%%%%%%%%%%%%%%%%%%%%%%%%%%%%%%%%%%%%%%%%%%%%%%%%%%%%%%%%%%%%%%%%%%%%%%%%%
\section{Periodic $\boldsymbol{(-\beta,l)}$-expansions}\label{sec:periodic}

Representations of numbers in Ito-Sadahiro numeration system (the case $l=-\frac{\beta}{\beta+1}$) from
the point of view of dynamical systems have been studied by
Frougny and Lai~\cite{ChiaraFrougny}. They have shown that if
$\beta$ is a Pisot number, then $\d(x)$ is eventually
periodic for any $x\in [\frac{-\beta}{\beta+1},\frac{1}{\beta+1})\cap\Q(\beta)$. In particular, their
result implies that for every Pisot number the reference strings $\d(l)$, $\d^*(r)$ in the admissibility
condition are eventually periodic.
Bases $\beta$, for which in the Ito-Sadahiro numeration system the reference string $\d(l)$ is eventually periodic
have been called Ito-Sadahiro numbers and by~\cite{ItoSadahiro} they are exactly the bases for which the Ito-Sadahiro $(-\beta)$-shift
is sofic. The result of Frougny and Lai thus implies that Pisot numbers are Ito-Sadahiro
numbers. This is an analogy to the statement for positive bases, namely that all Pisot numbers are Parry numbers.
Liao and Steiner~\cite{LiaoSteiner} show that surprisingly the notion of Ito-Sadahiro numbers and Parry numbers do not coincide.

In this section, we study the question of periodic $(-\beta,l)$-expansions.
We show that the statement on Ito-Sadahiro expansions can be simply extended to the case of
generalized $(-\beta)$-numeration defined on $[l,r)$, with
$l\in(-1,0]$ and $r=l+1$. Note that the proof, just as the one
in~\cite{ChiaraFrougny}, is only a slight modification of the
proof for positive bases given by Schmidt in~\cite{schmidt}.

\begin{thm}\label{thm:FL_generalised}
If $\beta$ is a Pisot number and $d$ is the $(-\beta,l)$-expansion for some $l\in(-1,0]$, then $\d(x)$ is eventually
periodic for any $x\in [l,r)\cap\Q(\beta)$.
\end{thm}

\pfz Let $\beta_2,\ldots,\beta_d$ be the algebraic conjugates of
$\beta=\beta_1$. Let $x\in[l,r)\cap\Q(\beta)$ be fixed. Such a
number obviously belongs to the set $\frac{1}{q}(\Z+\beta\Z+\cdots
+ \beta^{d-1}\Z)$ for some integer $q$. Denote $\d(x)=
x_1x_2x_3\cdots$ and write
$$
r_n=\frac{x_{n+1}}{-\beta}+\frac{x_{n+2}}{(-\beta)^2}+\ldots=(-\beta)^n\Big(x-\sum_{k=1}^n
x_k(-\beta)^{-k}\Big)\,.
$$
Note that since $\beta$ is an algebraic integer, we
have $\beta^k\in \Z+\beta\Z+\cdots + \beta^{d-1}\Z$, for every
non-negative integer $k$. This implies that
\begin{equation}\label{eq:rnmrizka}
r_n\in \frac{1}{q}\Big(\Z+\beta\Z+\cdots +
\beta^{d-1}\Z\Big)\quad\text{ for every } n\in\N.
\end{equation}

For $2\leq j\leq d$, we apply the isomorphism between $\Q(\beta)$ and $\Q(\beta_j)$ defined by
$$
z=c_0+c_1\beta + \cdots + c_{d-1}\beta^{d-1} \quad\mapsto\quad
z^{(j)}=c_0+c_1\beta^{(j)} + \cdots + c_{d-1}\big(\beta^{(j)}\big)^{d-1}
$$
to obtain
$$
r_n^{(j)}(x)=(-\beta_j)^n\Big(x^{(j)}-\sum_{k=1}^nx_k(-\beta_j)^{-k}\Big)\,.
$$
Denoting $\eta={\rm max}\{|\beta_j|, 2\leq j\leq d\}$, and since the digits $x_i$ of $\d(x)$ are
bounded by $\lceil\beta\rceil$, we find that
$$
|r_n^{(j)}|\leq\eta^nx^{(j)}+\lceil\beta\rceil\sum_{k=0}^{n-1}\eta^k\,.
$$
Since $\beta$ is a Pisot number, we have $\eta<1$, which implies that $|r_n^{(2)}|,\dots,|r_n^{(d)}|$ are bounded by a constant,
independent of $n\in\N$. Also $r_n^{(1)}=r_n$ is bounded, since $r_n=T^n(x)$, and thus $r_n\in[l,r)$, which implies
$|r_n|<1$.

Consider the lattice $L$ in $\R^d$ spanned by vectors
$$
\frac{1}{q}(1,1,\dots,1)\,,\quad
\frac{1}{q}\big(\beta_1,\beta_2,\dots,\beta_d\big)\,,\quad \dots,\quad
\frac{1}{q}\big(\beta_1^{d-1},\beta_2^{d-1},\dots,\beta_d^{d-1}\big)\,.
$$
From~\eqref{eq:rnmrizka}, it is obvious that the $d$-tuples
$$
R_1=(r_1,r_1^{(2)},\dots,r_1^{(d-1)}),\quad
R_2=(r_2,r_2^{(2)},\dots,r_2^{(d-1)}),\quad
\dots,\quad
R_n=(r_n,r_n^{(2)},\dots,r_n^{(d-1)}),\quad\dots
$$
belong to the lattice $L$ for all $n\in\N$, and we have derived that they are found within a bounded region of $\R^d$.
There are only finitely many lattice points in a bounded region, and hence one finds vectors $R_k$, $R_m$ for $k<m$ that coincide.
It follows that $r_{m}=r_k$ and $\d(x)$ is eventually periodic. \pfk

\begin{pozn}
Note that the above theorem does not speak about periodicity of $\d(l)$ if
$l\notin\Q(\beta)$. In fact, if $l\notin\Q(\beta)$, then the expansion
$\d(l)=l_1l_2l_3\cdots$ is necessarily non-periodic, even if $\beta$ is
a Pisot number.

On the other hand, the fact that $\d(l)$ is eventually
periodic does not predicate about the algebraic character of the
base $\beta$. It only says that $l\in\Q(\beta)$. However, assuming
that both reference strings appearing in the admissibility condition are eventually periodic
already implies that $\beta$ is an algebraic integer.
For, if $\d(l)$ and $\d^*(r)$ are eventually
periodic, we can derive an expression for the number $1={\scriptstyle  \bullet}\d^*(r) - {\scriptstyle  \bullet}\d(l)$ as a power series
in $(-\beta)^{-1}$ whose integer coefficients are arranged in an eventually periodic
order. Such an expression gives rise to a monic polynomial with integer coefficients having $\beta$ as a
root.
\end{pozn}

The following theorem is an almost reverse to
Theorem~\ref{thm:FL_generalised}. Assuming eventually periodic
$(-\beta,l)$-expansion of all rationals in the interval $[l,r)$,
we derive that $\beta$ must be Pisot or Salem. Here we would like to point out that
the original proof of Schmidt could not be adapted so easily because of the phenomena of degenerated intervals
mentioned in Section~\ref{sec:digitstrings}.

\begin{thm}\label{thm:nas_generalised}
Let $d$ be the $(-\beta,l)$-expansion. If any rational $x\in [l,r)$ has eventually periodic
$(-\beta,l)$-expansion, then $\beta$ is either Pisot or Salem
number.
\end{thm}

\pfz First realize that by assumption there exists a rational
number from $[l,r)$ of the form $\frac{1}{m}$, $m\in\Z$ with
eventually periodic expansion. It is easy to see that this gives
us a monic polynomial from $\Z[x]$ with $\beta$ as its root, hence
$\beta$ is an algebraic integer. It remains to show that all
conjugates of $\beta$ are in modulus smaller than or equal to $1$.

From Lemma~\ref{cl:pomocnyclaim} we derive that
there exists a non-zero digit $a$ such that for all $N\in\N$ we
can find a rational number $y$ satisfying
\begin{equation}\label{eq:pomocnyclaim2}
 \d(y) = a^Ny_1y_2y_3\cdots\,.
\end{equation}

Expression~\eqref{eq:pomocnyclaim2} can be rewritten
$$
y= \frac{a}{(-\beta)}+\cdots
+\frac{a}{(-\beta)^N}+\sum_{i=N+1}^\infty\frac{y_{i-N}}{(-\beta)^i}=a\frac{1-(-\beta)^{-N}}{\beta+1}+\sum_{i=1}^\infty\frac{y_{i}}{(-\beta)^{i+N}}
$$
and so
\begin{equation}\label{eq:rbeta}
y=\frac{a}{\beta+1}
+\frac{1}{(-\beta)^N}\left(\frac{-a}{\beta+1}+\sum_{i=1}^\infty\frac{y_i}{(-\beta)^i}\right)\,.
\end{equation}

As $y\in\Q$, by assumption, the infinite word
$y_{1}y_{2}y_3\cdots$ is eventually periodic and by summing a
geometric series, $\sum_{i=1}^\infty\frac{y_i}{(-\beta)^i}$ can be
rewritten as
$$
\sum_{i=1}^\infty \frac{y_i}{(-\beta)^i} = c_0+c_1\beta+\cdots
+c_{d-1}\beta^{d-1}\in\Q(\beta)\,,
$$
where $d$ is the degree of the algebraic integer $\beta$. In order
to prove the theorem by contradiction, assume that a conjugate
$\gamma\neq\beta$ is in modulus greater than 1. By application of
the isomorphism between $\Q(\beta)$ and $\Q(\gamma)$, we get
$$
c_0+c_1\gamma+\cdots +c_{d-1}\gamma^{d-1} = \sum_{i=1}^\infty
\frac{y_i}{(-\gamma)^i}\,,
$$
and thus from~\eqref{eq:rbeta} we get
\begin{equation}\label{eq:pf2}
y= \frac{a}{\gamma+1}
+\frac{1}{(-\gamma)^N}\left(\frac{-a}{\gamma+1}+\sum_{i=1}^\infty\frac{y_i}{(-\gamma)^i}\right)\,.
\end{equation}

Now denote $\eta = \max\{|\beta|^{-1},|\gamma|^{-1}\}<1$ and
realize that $\lceil\beta\rceil$ estimates the greatest digit in
modulus. Comparing~\eqref{eq:rbeta} and~\eqref{eq:pf2}, we obtain
\begin{equation}\label{eq:pf3}
0< \Big|\frac{a}{\beta+1} - \frac{a}{\gamma+1}\Big| \leq
\eta^{N}\left( \Big|\frac{a}{\beta+1}- \frac{a}{\gamma+1}\Big| +
2\lceil\beta\rceil \sum_{i=1}^\infty \eta^i
 %\sum_{i=1}^\infty |x_i|
%\Big|\frac{1}{(-\beta)^{i}}-\frac1{(-\gamma)^{i}}\Big|
\right)\,.
\end{equation}
Obviously, the value in the bracket on the right hand side
of~\eqref{eq:pf3} is a finite constant independent on $N$, and thus the
right hand side tends to zero with $N$ increasing to infinity. This leads to contradiction,
since on the left hand side of~\eqref{eq:pf3} we have a fixed
positive number.

%The proof of Theorem~\ref{thm:nas_generalised} will thus be
%completed by proving the above Lemma~\ref{cl:pomocnyclaim}.
%For that, consider a non-zero digit $a$ such that the corresponding interval $I_a\subset[c,c+1)$
%is non-degenerate. With the help of Remark~\ref{pozn:degenerovane_intervaly}, it is easy to see that
%such a digit always exists. Then, we take $f_a$ and another element $x\in I_a$, with expansions
%$d_{-\beta}^c(f_a)=a^\omega$ and $d_{-\beta}^c(x)=ax_2x_3\cdots$, such that
%$$
%ax_2x_3\cdots \prec_{\hbox{\tiny alt}} a^\omega\,.
%$$
%Such an $x$ can be always found, since $f_a$ is an inner point of the interval $I_a$.
%The digit string $a^{2k}ax_2x_3\cdots$, $k\in\N$, satisfies
%$$
%ax_2x_3\cdots \prec_{\hbox{\tiny alt}} a^{2k}ax_2x_3\cdots \prec_{\hbox{\tiny alt}} a^\omega\,,
%$$
%and thus us also admissible, i.e. it is an expansion of some $y\in[c,c+1)$. In particular,
%$y\in[x,f_a)$. It suffices now to chose a rational number $r\in [y,f_a)$. Its expansion has prefix
%$a^{2k}$. This completes the proof of Claim~\eqref{cl:pomocnyclaim} and thus the proof of Theorem~\ref{thm:nas_generalised}
%is also completed.
\pfk

\section*{Acknowledgements}

We acknowledge financial support by the Czech Science Foundation
grant 201/09/0584 and by the grants MSM6840770039 and LC06002 of
the Ministry of Education, Youth, and Sports of the Czech
Republic. The work was also partially supported by the CTU student
grant SGS. We are grateful to Wolfgang Steiner for useful disscussions and pointing
out Example~\ref{ex:wolfik} to us.

%%%%%%%%%%%%%%%%%%%%%%%%%%%%%%%%%%%%%%%%%%%%%%%%%%%%%%%%%%%%%%%%%%%%%%%%%%

%%%%%%%%%%%%%%%%%%%%%%%%%%%%%%%%%%%%%%%%%%%%%%%%%%%%%%%%%%%%%%%%%%%%%%%%%%
\end{document}